\documentclass[11pt]{article}

\usepackage[english]{babel}
\usepackage[letterpaper,top=3cm,bottom=3cm,left=3.1cm,right=3.1cm,marginparwidth=1.2cm]{geometry}
\usepackage{amsfonts}
\usepackage[dvipsnames]{xcolor}
\usepackage{amsmath}
\usepackage{graphicx}
\usepackage[colorlinks=true,allcolors=blue]{hyperref}
\usepackage{multirow}
\usepackage{tikz}
\usepackage{hhline}
\usepackage{float}
\usepackage{todonotes}
\usepackage{titlesec}
\usepackage{array}
\usepackage{tabularx}
\usepackage{adjustbox}
\usepackage[referable]{threeparttablex}
\usepackage{footnote}
\usepackage{hyperref}
\usepackage[skip=9pt, indent=20pt]{parskip}

\usepackage{authblk}

\newcommand{\ti}[1]{\textit{#1}}

\newcommand{\RR}{\mathbb{R}}
\newcommand{\CC}{\mathbb{C}}

\newcommand{\MM}{\mathcal{M}}

\newcommand{\VV}{\mathcal{V}}

\newcolumntype{M}[1]{>{\centering\arraybackslash}m{#1}}

\usepackage{amsthm}
\newtheorem{defn0}{Definition}[section]
\newtheorem{prop0}[defn0]{Proposition}
\newtheorem{thm0}[defn0]{Theorem}
\newtheorem{lemma0}[defn0]{Lemma}
\newtheorem{corollary0}[defn0]{Corollary}
\newtheorem{example0}[defn0]{Example}
\newtheorem{conjecture0}[defn0]{Conjecture}
\newtheorem{notation0}[defn0]{Notation}
\newtheorem{remark0}[defn0]{Remark}
\newtheorem{assumption0}[defn0]{$d$-claw tree hypothesis}
\newtheorem{problem0}[defn0]{Problem}

\newenvironment{defi}{\begin{defn0} \rm}{\end{defn0}}

\newenvironment{thm}{\begin{thm0}}{\end{thm0}}
\newenvironment{lema}{\begin{lemma0}}{\end{lemma0}}
\newenvironment{cor}{\begin{corollary0}}{\end{corollary0}}
\newenvironment{example}{\begin{example0} \rm}{\end{example0}}
\newenvironment{rk}{\begin{remark0} \rm}{\end{remark0}}
\newenvironment{conjecture}{\begin{conjecture0}\rm}{\end{conjecture0}}

\title{\textbf{Computing algebraic degrees \\ of phylogenetic varieties}}

\author[1]{Luis David Garc\'{\i}a Puente \thanks{ lgarciapuente@coloradocollege.edu}}
\author[2]{Marina Garrote-L\'opez \thanks{mgarrote@math.ubc.ca}}
\author[2,3]{Elima Shehu\thanks{elima.shehu@uni-osnabrueck.de}}
\affil[1]{{\footnotesize Department of Mathematics and Computer Science, Colorado College, Colorado Springs, CO, USA}}
\affil[2]{{\footnotesize Max Planck Institute for Mathematics in the Sciences, Leipzig, Germany}}
\affil[3]{{\footnotesize Osnabrück University, Osnabrück, Germany}}
\date{}

\begin{document}
\maketitle
\vspace{-0.4cm}

\begin{abstract}
Phylogenetic varieties are algebraic varieties specified by a statistical model describing the evolution of biological sequences along a tree. 
Its understanding is an important problem in algebraic statistics, particularly in the context of phylogeny reconstruction.
In the broader area of algebra statistics, there have been important theoretical advances in computing certain invariants associated with algebraic varieties arising in applications. Beyond the dimension and degree of a variety, one is interested in computing other algebraic degrees, such as the maximum likelihood degree and the Euclidean distance degree. Despite these efforts, the current literature lacks explicit computations of these invariants for the particular case of phylogenetic varieties. In our work, we fill this gap by computing these invariants for phylogenetic varieties arising from the simplest group-based models of nucleotide substitution Cavender-Farris-Neyman model, Jukes-Cantor model, Kimura 2-parameter model, and the Kimura 3-parameter model on small phylogenetic trees with at most 5 leaves. 
\end{abstract}

\section{Introduction}

One of the main goals of phylogenetics is to reconstruct the evolutionary history of a group of different biological entities, such as genes, bacterial strains or species. In phylogenetics, we model these evolutionary relationships with a phylogenetic tree, where the leaves represent the extant entities of interest and the internal nodes represent their extinct common ancestors. After some mild hypothesis, to model the substitution of nucleotides in an evolutionary process we adopt a parametric statistical model, that is, we model the substitution of nucleotides along a phylogenetic tree as a Markov process on the tree. This allows us to define a parameterization map that describes the joint distributions at the leaves of the tree in terms of the parameters of the model.

The resulting polynomial map $\phi^{\MM}_{T}:\mathbb R^{d} \rightarrow \mathbb R^{m}$ associated to a tree $T$ and a model $\MM$ maps any $d$-tuple of parameters to a distribution vector of the $m$ possible observations at the leaves. 
The Zariski closure of the image of this map denoted $\overline{Im(\phi^{\MM}_{T}})$, is the \emph{phylogenetic variety} associated to $T$ and $\MM$. For certain evolutionary models, this variety is a projective toric variety. 

In applications, one is interested in finding the phylogenetic tree and the parameters that best explain some given evolutionary data. Geometrically, this can be interpreted as, given a data point compute the closest point on the varieties associated to the different possible trees. It is known that, for complex varieties, the number of complex non-singular critical points of the squared Euclidean distance to a generic point outside the variety is constant in a dense open subset, \cite{draisma2015}. This number, known as the \emph{Euclidean distance degree} (ED degree) of the variety, measures the algebraic complexity of the variety providing an upper bound for the number of real solutions. 
Another approach to recover the tree and parameters that best explain some data is by maximizing the likelihood of the observed distribution. That is, the maximum likelihood estimation is a similar optimization problem where one tries to find the model parameters that best explain the data. The \emph{maximum likelihood degree} (ML degree) is the number of complex solutions to the critical equations of this optimization problem, which is also shown to be constant \cite{HKS05, catanese2006maximum}. 

Computing the ML degree of phylogenetic varieties is a classical problem, and some degrees are already known for certain trees and models, see Section \ref{sec:MLE}. On the other hand, the study of ED degrees of these varieties is an emerging and understudied problem. To the best of our knowledge, only the ED degree of one tree of $4$ leaves with a specific evolutionary model has been studied so far (see \cite{casfergarrote2020}).

In this work, we aim to extend the computations of ED and ML degrees for some group-based models on small phylogenetic trees with at most 5 leaves. 
Our approach is fully computational and we use both symbolic tools as well as tools from numerical algebraic geometry that allow us to compute lower bounds for the ED and ML degrees.
The main contribution of this paper is Theorem \ref{main}, where we establish a lower bound for the algebraic degrees of trees with up to 5 leaves and group-based models. To obtain these values, we also require Lemma \ref{lema:cardinallity}, where we present the cardinalities of the generic fibers of the parametrization map $\phi^{\MM}_{T}$ for the group-based models.

The paper is organized as follows, in Section \ref{sec:phylo} we introduce the basic notions on phylogenetic trees, phylogenetic varieties, and group-based models. In addition, we present the Fourier transform, which allows the parametrization of phylogenetic varieties through a monomial map.
In Section \ref{sec:degrees} we describe in more detail the ML degree and the ED degree for phylogenetic varieties. Section \ref{sec:comp_results} is dedicated to presenting the computational results we have obtained using symbolic and numerical methods, and these findings are subsequently discussed in Section \ref{sec:discussion}.

\section{Phylogenetic models}\label{sec:phylo}

In this section, we introduce phylogenetic models and the algebraic varieties associated to them. We focus on group-based phylogenetic models and introduce a change of coordinates called the discrete Fourier transformation, in which the parameterization of the phylogenetic varieties
becomes monomial. The reader is referred to 
\cite{Felsenstein2003, semple2003phylogenetics, ASCB2005, Sturmfels2005}
for a more detailed introduction to the topic.

Evolutionary relationships between a set of biological entities, or \textit{taxa}, are commonly depicted in the form of {phylogenetic trees}. A \textit{phylogenetic tree} $T$ is a connected acyclic graph whose leaves are in one-to-one correspondence with a set of labels. That is, the leaves $L(T)$ of a phylogenetic tree $T$ represent a set of current taxa, the inner nodes $\text{In}(T)$ correspond to common ancestors, and the edges $E(T)$ represent the evolutionary relationships among them. For simplicity, we assume leaves are in correspondence with the set $\{1,2,\ldots, n\}$ where $n$ is the number of leaves.
We choose an internal vertex of $T$ as the root $r$ of $T$, {which induces an orientation on the set of edges, from the root to the leaves}. 

The substitution of nucleotides in an evolutionary process is modeled by adopting a parametric statistical model as follows: associate a random variable $X_v$ taking values on a set of states $\Sigma:=\{s_1,\ldots,s_k\}$\footnote{$\Sigma$ is usually assumed to be the set of nucleotides $\{\tt A,\tt C,\tt G,\tt T \}$ or the binary set $\{0,1\}$ that can be identified with the two groups of nucleotide bases purines and pyrimidines.} 
at each node $v$ of the tree. Also, consider as parameters a distribution $\pi = (\pi_{1}, \ldots, \pi_{k})$ at the root $r$ of the tree, where $p_{i} = Prob(X_r=s_i)$ and such that $\sum_i\pi_i = 1$; and a $k\times k$ transition matrix $M_e$ at each {(oriented)} edge $e:u\to v$ of $T$ given by $(M_e)_{ij} = Prob(X_v=s_j|X_u=s_i)$. The transition matrices are \textit{stochastic} (or \textit{Markov}) matrices, that is, all its entries are non-negative and its rows sum up to $1$.

Assuming a hidden Markov process on a phylogenetic tree, the probability of observing a particular set of nucleotides at the leaves of the tree can be computed as a polynomial in terms of the model parameters as follows,
\begin{equation}\label{eq:jointDistribution}
p_{x_1,\ldots,x_n}=\sum_{\substack{x_r,x_v\in\Sigma\\ {\scriptsize v\in \text{In}(T)}}}\pi_{x_r}\prod_{e\in E(T)} M_e\left(x_{u},x_{v}\right),
\end{equation}
where $x_r$ is the state of the root, $x_u$ is the state of the parent node $u$ of $e$ , and $x_v$ is the state of the child node $v$ of $e$. If $e$ is a terminal edge ending at leaf $i$ then $x_v=x_i$.

We consider the polynomial map $\phi_T^\MM:\mathbb{R}^d\to\mathbb{R}^{m}$ associated to a tree $T$ and model $\MM$ that maps any $d$-tuple to a distribution vector of the $m:=k^n$ possible observations at the leaves,
\begin{equation}\label{joint-dist}
\begin{array}{rccc}
\phi_T^{\mathcal{M}}: & \Theta\subset\mathbb{R}^d & \to &\bigtriangleup^{m-1}\subset \mathbb{R}^{m} \\
& \theta = \left(\theta_1,\ldots,\theta_d\right) &\mapsto & p=(p_{ \tt s_1 s_1 \dots s_1},p_{ \tt s_1s_1 \dots s_2},p_{ \tt s_1 s_1 \dots s_3}, \dots, p_{ \tt s_k s_k \dots s_k}),
\end{array}
\end{equation}
where $d$ is the number of free substitution parameters (that is, free entries of the distribution at the root $\pi$ and the transition matrices $M_e$) according to the model assumed and $\Theta$ denotes the space of stochastic parameters. 
The coordinates $p_{x_1,\ldots,x_n}$ correspond to the joint distribution of characters at the leaves of $T$ given by a hidden Markov process on $T$ governed by these parameters.
Our assumption that the entries of $\pi$ and rows of $M_e$ sum to $1$ implies $\sum p_{x_1,\ldots,x_n} =1$.
Ignoring the stochastic restrictions on the parameter space, we may regard $\phi_T^\MM:\mathbb{C}^d\to\mathbb{C}^{m}$ as a complex polynomial map. The Zariski closure of the image of such complex map denoted $\VV_T^{\MM} = \overline{\phi_T^{\MM}(\mathbb{C}^d)}\in\mathbb{C}^{m}$, is an algebraic variety known as the \textit{phylogenetic variety} associated to $T$ and $\MM$.

\newpage
\paragraph{Group-based models}\label{sec:models}

Nucleotide substitution models $\MM$ are specified by certain restrictions, which may arise from biochemical properties, on the distribution at the root and the transition matrices. The group-based models are certain substitution models in which the set $\Sigma$ is given the additional structure of an abelian group. That is, a \ti{group-based model} relative to an abelian group $G$ is a substitution model whose matrices can be defined in terms of a function $f^e: G\to \RR$ such that $M^e(g,h)=f^e(g-h)$, for any $g,h\in G$ after the identification of $\Sigma$ with $G$. 

In our work, we consider the Cavender-Farris-Neyman (CFN) model 
where $\Sigma=\{0,1\}$ is identified with the group $\mathbb{Z}/2\mathbb{Z}$ and the \textit{$3$-parameter Kimura} (K81) model \cite{Kimura1981} that identifies $\Sigma=\{\tt A,\tt C,\tt G,\tt T \}$ with the group $\mathbb{Z}/2\mathbb{Z} \times \mathbb{Z}/2\mathbb{Z}$.
Both models assume a uniform distribution at the root of the tree and respective transition matrices have the form
 $$M_e=
		\left(
		\begin{array}{cc}
		a & b \\
		b & a \\
		\end{array}
		\right), \qquad
		M_e=
		\left(
		\begin{array}{cccc}
		a & b & c & d \\
		b & a & d & c \\
		c & d & a & b \\
		d & c & b & a \\
		\end{array}
		\right).
		$$
These two models are referred to as \textit{general group-based models} and they identify the set of states with a group $G$. For more details on the group structure of these models, the reader is referred to \cite{sullivant2018algebraic}. 
We also consider two submodels of the K81 model, the \textit{Jukes-Cantor} ($JC69$) model \cite{JC69} and the \textit{$2$-parameter Kimura} (K80) model \cite{Kimura1980}. Transition matrices for these models can be obtained from K81 matrices by identifying parameters, that is, the K80 model and the JC69 model assume $c=d$ and $b=c=d$, respectively.

The parametrization map $\phi_T^{\mathcal{M}}$ in general is not one-to-one. 
Understanding its generic fibers provides important computational and biological insight. From the perspective of reconstructing the evolutionary history, knowing the size of a generic fiber gives the number of possible evolutionary scenarios for the same observed data. In particular, when this number is small, this invariant is an indicator that the evolutionary history is relatively robust when reconstructed.

In what follows we present the cardinality of such fibers, this corresponds to the number of parameters of the model that are mapped by $\phi_T^{\mathcal{M}}$ to the same probability distribution at the leaves of the tree.

\begin{lema}\label{lema:cardinallity}
 The cardinalities $\delta_T^\MM$ of the generic fibers of $\phi_T^{\mathcal{M}}$ for the previously described group-based models are 
 \begin{equation}\label{eq:fibers}
 \delta_T^{CFN} = 2^N, \quad \delta_T^{JC} = 1, \quad \delta_T^{K80} = 2^N \quad\mbox{and}\quad \delta_T^{K81} = 4^N
 \end{equation}
 where $N$ is the number of interior nodes of $T$.
\end{lema}

\begin{proof}
The result is already known for any general group-based model, see Proposition 2.15 in \cite{casfermich2015} where the authors prove that the cardinality of the generic fiber of $\phi_T^{\mathcal{M}}$ is 
 $\lvert G \rvert^N,$
where $G$ is the group corresponding to the model $\MM$.

For the other models, the result follows from the fact that certain permutations of the rows and columns of the Markov matrices $\left\{M_e\right\}_{e\in E(T)}$ can be applied without affecting the joint distribution $p$ (see \cite{Allman2003} or \cite{Chang96}). 
That is, consider two distributions $p = \phi_T\left(\pi, \left\{M_e\right\}_{e\in E(T)}\right)$ and $p' = \phi_{T'}\left(\pi', \left\{M'_e\right\}_{e\in E(T)}\right)$ such that $p = p'$. According to \cite[Definition $3.5$ and Proposition $3.7$]{allman2004b}, then $T=T'$ and there exists a set of permutations matrices $P_{v_1}, \ldots, P_{v_N}$ for each interior node $v_i\in Int(T)$, such that $\pi' = \pi P_r^T$ and $M'_e = P_uM_eP_v^T$ for $e:u\to v$, with $P_v = Id$ if $v$ is a leaf of the tree. This implies that there are at most $(|\Sigma|!)^N$ set of parameters that map to the same distribution.

However, if we fix a particular model, the number of permutations that one can consider such that the structure of the transition matrices is preserved is much smaller than $|\Sigma|!$.
We start considering the K81 model and a tree $T$. 
Let $e:u\to v$ be a terminal edge of $T$, that is, $v$ is a leaf, and $M_e$ is the transition matrix associated with $e$. Consider a permutation matrix $P_u$ associated to the vertex $u$, then we can replace $M_e$ with the matrix $P_uM_e$.
It can be easily checked that, if $M_e$ is a K81 matrix, then $P_uM_eId$ is a K81 matrix if and only if the permutations $P_u$ is one of the following matrices:
\begin{equation}\label{eq:permu}
 \Gamma_1 = Id, \ 
		\Gamma_2 = \left(
		\begin{array}{cccc}
		0 & 1 & 0 & 0 \\
		1 & 0 & 0 & 0 \\
		0 & 0 & 0 & 1 \\
		0 & 0 & 1 & 0 \\
		\end{array}
		\right), \ 
		\Gamma_3 = \left(
		\begin{array}{cccc}
		0 & 0 & 1 & 0 \\
		0 & 0 & 0 & 1 \\
		1 & 0 & 0 & 0 \\
		0 & 1 & 0 & 0 \\
		\end{array}
		\right) \ 
		\Gamma_4 = \left(
		\begin{array}{cccc}
		0 & 0 & 0 & 1 \\
		0 & 0 & 1 & 0 \\
		0 & 1 & 0 & 0 \\
		1 & 0 & 0 & 0 \\
		\end{array}
		\right).
\end{equation} 

Consider now and edge $e:u\to v$ such that $v$ is the parent node of a leaf of $T$. Let $P_u$ and $P_v$ be two permutation matrices associated to $u$ and $v$. As shown above, to keep the structure of the transition matrices, $P_v$ is one of the matrices $\{\Gamma_1,\Gamma_2,\Gamma_3,\Gamma_4\}$. Then, it is straightforward to see that $P_uM_eP_v^T$ is a K81 matrix if and only if $P_u$ is also one of the permutation matrices in \eqref{eq:permu}. This same argument can be applied recursively to all edges of the tree, showing that the only permutation matrices $P_u$ and $P_v$ that can be considered for $P_uM_eP_v^T$ to preserve the K81 structure are the matrices in $\{\Gamma_1,\Gamma_2,\Gamma_3,\Gamma_4\}$. This implies that the number of parameters mapping to the same distribution by $\phi_T^{\mathcal{M}}$ is $4^N$.
Recursively along all edges of $T$ one can see that all chosen $P_u$ for any inner node $u$ should be in $\{\Gamma_1,\Gamma_2,\Gamma_3,\Gamma_4\}$ so that $P_uM_ePv^T$ are K81 matrices.

A similar argument can be applied to the K80 model, but in this case there are only two matrices, $\Gamma_1$ and $\Gamma_2$ that preserve the structure of the transition matrices, which implies that the cardinality of the fibers of $\phi_T^{K80}$ is $2^N$. Finally, $\delta_T^{JC69}=1$ follows from the fact that the only permutation $P_u$ that preserves the structure of the JC69 matrices is the identity matrix.
\end{proof}

For more insight on group-based models and their corresponding phylogenetic varieties, we refer to \cite{Sturmfels2005}, \cite{Michalek2011}, and \cite{casfermich2015}.

\paragraph{Fourier transform}\label{sec:fourier}
The discrete Fourier transform is a powerful tool to analyse the properties of group-based models and the pattern distributions arising from them. This idea was introduced by \cite{hendy89} and \cite{hendyPenny89} and was further explored by \cite{SZEKELY1993200}.
The discrete Fourier transform provides a linear change of coordinates that reveals the irreducible variety of distributions of a group-based model $\VV_T^{\MM}$ to be a toric variety. The following results provide an explicit description of the parameterization of these varieties after this linear change of coordinates.

\begin{thm}[\cite{Evans1993}]
Let $p$ be the joint distribution of a group-based model for a phylogenetic tree $T$ obtained as in \eqref{eq:jointDistribution}. Then the Fourier transform of $p$ has the form
 \begin{equation}\label{eq:fourier_param}
	q_{y_{1}\ldots y_{n}} =
	\begin{cases}
	\prod_{e\in E(T)} x_{k}^e & \text{if $y_n=y_1+\cdots+y_{n-1}$,}
	\\
	0 &\text{otherwise}
	\end{cases}
	\end{equation}
 where $k=\sum_{i\in de(e)}y_i$, and $de(e)$ is the set of all leaves $l$ descendant of $e$\footnote{A leaf $l$ is a descendant of an edge $e$ if there is a directed path from $e$ to $l$.} and $x^e$ are the \textit{Fourier parameters} associated to edge $e$.
\end{thm}

{Fourier parameters} are the Fourier transforms of the transition functions $f^e: G\to\RR$, which can be shown to be equivalent to the eigenvalues of transition matrices $M_e$. 
For simplicity, we assume $x_0^e,\ldots,x_l^e$ are the $l+1$ different eigenvalues of $M_e$, where $l$ coincides with the number of free parameters in $M_e$. 
As a consequence of the previous theorem, we can consider a map $\varphi_T^{\mathcal{M}}:\CC^d \to \CC^m$ parameterizing phylogenetic varieties from Fourier parameters to Fourier coordinates. Since the map $\varphi_T^{\mathcal{M}}$ is homogeneous we can consider the Zariski closure of the image of such map as a projective variety in $\mathbb{P}^{m-1}:$
\begin{equation}\label{eq:jointDistFourier}
\begin{array}{rccc}
\varphi_T^{\mathcal{M}}: & \mathbb{P}^l \times \cdots \times \mathbb{P}^l & \to & \mathbb{P}^{4^n-1} \\
& \left([x_0^{e_1}:\dotsm :x_l^{e_1}], \ldots, [x_0^{e_E}:\dotsm :x_l^{e_E}]\right) & \mapsto & q=(q_{ \tt A \dots A}, \dots, q_{ \tt T \dots T}),
\end{array}
\end{equation} where $E$ denotes the number of edges $E(T)$ of $T$.

\begin{rk}
 The variety $\overline{\VV_T^\MM} = \overline{\text{Im }\varphi_T^{\mathcal{M}}}$ is the projectivization of the affine phylogenetic variety $\VV_T^\MM$. That is, the affine variety $\VV_T^\MM$ corresponds to the affine chart given by $x_0^{e_1} = \dots = x_0^{e_E} = 1$.
\end{rk}

In Fourier coordinates, $\overline{\VV_T^\MM}$ 
is a complex toric variety that admits a monomial parameterization $\varphi_T^\MM$. This monomial map can be described by a $d\times m$ matrix $A$ whose columns are the exponent vectors of the monomials parameterizing each Fourier coordinate.

\begin{example} Consider a $3$-leaf binary tree $T_3$ with leaves $[3]$ evolving under the CFN model. Consider the Fourier parameters $(x_0^{e_1},x_1^{e_1}),\ (x_0^{e_2},x_1^{e_2}),\ (x_0^{e_3},x_1^{e_3})$, where each $e_i$ denotes the edge pointing to the leaf $i$. Then the nonzero Fourier coordinates are
$$q_{\tt 000}= x_0^{e_1}x_0^{e_2}x_0^{e_3}, \quad q_{\tt 011}=  x_0^{e_1}x_1^{e_2}x_1^{e_3}, \quad q_{\tt 101}=  x_1^{e_1}x_0^{e_2}x_1^{e_3}, \quad \text{and} \quad q_{\tt 110}=  x_1^{e_1}x_1^{e_2}x_0^{e_3}.$$
The variety $\overline{\VV_{T_3}^{CFN}}$ can be parametrized via the matrix of exponents 
$$A = 
\bordermatrix{ & {\footnotesize q_{000}} & {q_{000}} & {\footnotesize q_{000}} & {q_{000}} \cr
 x_0^{e_1} & 1& 1& 0& 0 \cr
 x_1^{e_1} & 0& 0& 1& 1 \cr
 x_0^{e_2} & 1& 0& 1& 0 \cr
 x_1^{e_2} & 0& 1& 0& 1 \cr
 x_0^{e_3} & 1& 0& 0& 1 \cr
 x_1^{e_4} & 0& 1& 1& 0
 }. 
$$ 
The matrices $A$ can be constructed with the function \emph{phyloToricAMatrix} from the Macaulay2 package \texttt{PhylogeneticTrees} \cite{phylotreesM2}.
\end{example}
Since the discrete Fourier transform is a linear change of coordinates, it follows from Lemma \ref{lema:cardinallity} that the cardinality of the fibers of the parametrization map on Fourier parameters $\varphi_T^{\mathcal{M}}$ is preserved.

\begin{cor}
 The cardinalities of the generic fibers of the map $\varphi_T^{\mathcal{M}}$ for the models CFN, JC69, K80, K81 models are $2^N$, $1$, $2^N$ and $4^N$ respectively.
\end{cor}

We conclude the section with the following remark.
\begin{rk}\label{rk:singular}
The singular points of the varieties $\overline{\VV_T}$ and ${\VV_T}$ under group-based models model are those that are image of some null parameter. In other words, $\varphi_T^{\mathcal{M}}(x^{e_1},\dots, x^{e_N})$ is a singular point of the variety if and only if $x_i^{e_j}=0$ for some $i$ and edge $e_j$. We refer to the works \cite{casfer2008} and \cite{casfermich2015} for details.
\end{rk}

\section{Algebraic degrees for phylogenetic models}\label{sec:degrees}

In this section, we discuss an important problem in mathematical phylogenetics: given some data $u$ find a distribution $p$ in $\mathcal{V}_T$ that makes $u$ most likely. Such distribution $p$ may depend on the distance measure chosen. In this work, we focus on two classical optimization problems in algebraic statistics, the maximum likelihood estimate and the closest point on a variety in the Euclidean distance. 
First, we introduce the \textit{Maximum likelihood estimator} problem and then the problem of minimizing the Euclidean distance over an algebraic variety.

\paragraph{Maximum likelihood estimator}\label{sec:MLE}

Maximum likelihood estimation is a standard approach to parameter estimation and a fundamental computational task in statistics. In the case of phylogenetics, given observed data $u$, a phylogenetic tree $T$, and a nucleotide substitution model $\MM$, the maximum likelihood estimate is the set of parameters $\theta=(\theta_1,\ldots,\theta_d)$ that is most likely to have produced the data $u$. We assume data is given by a non-negative integer vector $u$ in $\mathbb{N}^n$, where $u_i$ is the number of times the $i$-th event is observed. Then, the problem of \textit{maximum likelihood estimation} aims to find the parameters $\theta$ that maximizes the log-likelihood function
\begin{equation}\label{eq: lu} \ell_u(\theta) = \sum_{i=1}^{m} u_i\ log(p_i),\end{equation}
where $p = (p_1,\ldots,p_m) = \phi_T^{\mathcal{M}} (\theta).$
Local and global optimizers of the log-likelihood function are solutions of the \textit{critical equations} of $\ell_u(\theta)$, which are given by 
\begin{equation}\label{eq:critEqs}
 \frac{\partial \ell_u}{\partial \theta_1} = \dotsm = \frac{\partial \ell_u}{\partial \theta_d} = 0 
\end{equation}
For generic data $u$, the number of complex solutions $\theta\in\mathbb{C}^d$ of the critical equations $\frac{\partial \ell_u}{\partial \theta_i} =0$ such that $\phi_T^{\mathcal{M}}(\theta)$ is a smooth point in $\VV_T$ is finite and constant (see \cite{HKS05, catanese2006maximum}):

\begin{defi}
The \textit{maximum likelihood degree} (ML degree) of a tree $T$ and model $\MM$ is the number of complex solutions of the system of critical equations of $\ell_u(\theta)$, for generic data $u$, divided by $\delta_T^\MM$, the cardinality of the fibers of $\phi_T^{\mathcal{M}}$(see Remark \ref{lema:cardinallity}).
\end{defi}

ML degrees of phylogenetic varieties have been studied extensively, and they are known for some specific trees and models. For instance, the ML degree for the $3$-leaf star tree evolving under the Jukes Cantor model is $23$ and can be found in \cite{HKS05}.\linebreak Known ML degrees of phylogenetic varieties are compiled in \cite{CGS} and the companion website:
    \href{https://www.coloradocollege.edu/aapps/ldg/small-trees/}{www.coloradocollege.edu/aapps/ldg/small-trees/}.
In section \ref{sec:comp_results}, we extend these computations and present new ML degrees that have been obtained computationally. 

\paragraph{Minimizing the Euclidean distance}\label{sec:ED}

A different approach to finding parameters that better explain a data point $u$ is to minimize the Euclidean distance function from $u$ to a given algebraic variety $\VV_T^\MM$. In this case, we assume that the data $u \in \mathbb{R}^m$ is a distribution. The aim is to minimize the squared Euclidean distance function from $u$ to the variety $\VV_T^\MM$. In order to accomplish this, we consider the Fourier parameterization of the variety introduced in \eqref{eq:jointDistFourier}. Since the Fourier transformation is an orthogonal linear change of coordinates, the Euclidean distance between distributions can be computed using these new coordinates (see Remark 2.5 in \cite{casfergarrote2020} for more details). Then we consider the following optimization problem
\begin{equation}
\min_{x} d_{u}(x) := \min_{q\in\overline{\VV_T^\MM}} \sum_{i=1}^m (u_i - q_i)^2,
\label{eq:du}
\end{equation}
where $x = \left([x_0^{e_1}:\dotsm :x_l^{e_1}], \ldots, [x_0^{e_E}:\dotsm :x_l^{e_E}]\right)$ are the Fourier parameters and $q = (q_1,\ldots, q_m) = \varphi_T^{\mathcal{M}} (x)$ the Fourier coordinates.
The critical points of this problem are the solutions of the system 
\begin{equation}\label{eq:critEqsED}
\frac{\partial d_u}{\partial x_1} = \dots = \frac{\partial d_u}{\partial x_d} =0. 
\end{equation} 

\begin{defi}
the \textit{Euclidean distance (ED) degree} of $\VV_T^{\MM}$ is the number of complex solutions of the critical equations $\frac{\partial d_u}{\partial x_i} =0$ at which the Jacobian of $\varphi_T^{\mathcal{M}}(x)$ has maximal rank, divided by $\delta_T^\MM$ (for generic data $u$). 
\end{defi}

\begin{rk}\label{rk:EDD}
    That is, for generic data $u$, the number of complex solutions $x\in\mathbb{C}^d$ of the critical equations $\frac{\partial d_u}{\partial x_i} =0$ such that $\varphi_T^{\mathcal{M}}(x)$ is a smooth point in $\VV_T$ is constant, see \cite{draisma2015}. 
\end{rk}

In this work, we also consider a more general problem, that is, given a vector $\lambda = (\lambda_1,\ldots,\lambda_m)$, consider the minimization problem 
\begin{equation}
\min_{x} d_{u,\lambda}(x) = \min_{q\in\overline{\VV_T^\MM}} \sum_{i=1}^m \lambda_i(u_i - q_i)^2.
\label{eq:gen_du}
\end{equation}
If the weight vector $\lambda$ is generic, then the number of complex critical points of \eqref{eq:gen_du}, that is, the solutions of 
\begin{equation}\label{eq:critEqsgED}
\frac{\partial d_{u,\lambda}}{\partial x_1} = \dots = \frac{\partial d_{u,\lambda}}{\partial x_d} =0, 
\end{equation} is independent of $\lambda$ and is called the \textit{generic ED degree} (gED degree). 
Thus, the generic ED degree can be computed in terms of the Chern-Mather volumes of all faces of the polytope $P=\text{conv}(A)$where $A$ is the matrix defining the monomial map previously described. For the explicit expression of the gED degree for toric varieties, the reader is referred to \cite[Theorem 1.1]{Helmer_Sturmfels_2018}. 
Studying the structure of the A matrices for different trees or models might be a promising approach to obtain a general expression for the gED. However, we have not observed any discernible pattern in the $A$ matrices, which makes it difficult to conjecture any formula.

\begin{rk}
According to Theorem 6.11 in \cite{draisma2015}, the gED degree of $\overline{\VV_T^\MM}$ is an upper bound of both the ED degree of $\overline{\VV_T^\MM}$ and the ED degree of ${\VV_T^\MM}$.
\end{rk}

{The study of the optimization problem \eqref{eq:du}, i.e. minimizing the Euclidean distance from a data point to a phylogenetic variety, as well as the calculation of ED degrees for these varieties is still recent. As far as we are aware, only the ED degree for the 4-leaf binary tree evolving under the JC69 model is previously known, see \cite[Lemma 5.1]{casfergarrote2020}.}

\section{Computational results}\label{sec:comp_results}

In this section, we present the computational results that we obtained for the algebraic degrees for phylogenetic trees up to $5$ leaves and the group-based models previously introduced. 
More precisely, we computed the Euclidean distance degree and the generic Euclidean distance degrees of the affine varieties $\VV_T^\MM$ and its projectivization $\overline{\VV_T^\MM}$, that is 
$$\mbox{ED deg of }{\VV_T^\MM} \qquad  \mbox{gED deg of }{\VV_T^\MM}  \qquad  \mbox{ED deg of }\overline{\VV_T^\MM} \qquad \mbox{gED deg of }\overline{\VV_T^\MM}$$ 
for the trees $T$ and group-based models $\MM$ detailed below.
We also computed the ML degree for the affine varieties $\VV_T^\MM$.

In what follows, we first discuss the computational approaches that we used, then we present the obtained degrees and finally the execution times for such computations. The implementation of all methods discussed in this contribution can be found at \linebreak
    \href{https://github.com/marinagarrote/Computing-algebraic-degrees-of-phylogenetic-varieties}{www.github.com/marinagarrote/Computing-algebraic-degrees-of-phylogenetic-varieties}.

\subsection{Computational approaches}\label{sec:computations}

In order to compute the described algebraic degrees we used both, symbolic and numerical approaches. In what follows we present all these different approaches, mention their strengths and limitations, and specify which have been used to calculate the different types of degrees.

\paragraph{Symbolic methods}

We have followed two different symbolic approaches to compute the ED degrees. 
First, we follow \cite{draisma2015} to compute the ED degree of the varieties by computing the degree of the ideal $I$ defined by the critical equations described in \eqref{eq:critEqsED} and \eqref{eq:critEqsgED}. However, this computation can not be done straightforwardly, as one first needs to compute the saturation ideal $I:\langle x_1\cdots x_n\rangle^\infty$, to dismiss the singular part of the variety (see Remarks \ref{rk:singular} and \ref{rk:EDD}). 
This computation is done by computing a Gröbner basis and it is in general hard and very time-consuming. For this reason, in this work, we have performed the computations of these saturation ideals over finite fields $\mathbb{Z}/p\mathbb{Z}$ for large primes $p$, as we are only interested in the degree of the variety, and not the solutions of the optimization problem. As we mention in the discussion below we obtain consistent results with a faster time of execution.
We have used the computer algebra system \texttt{OSCAR} \cite{OSCAR, OSCAR-book}, written in \texttt{Julia} \cite{julia} to compute the saturation ideal and its degree. The saturation ideal is computed using the \texttt{msolve} library \cite{msolve} which uses a new approach to compute the Gröbner basis.

On the other hand, we have used the Macaulay2 package \texttt{toricInvariants} (see \cite{Helmer_Sturmfels_2018}) to compute the generic ED degree of the projective toric varieties parameterized by \eqref{eq:fourier_param}. This computation is based on computing the Chern-Mather volumes of all faces of the polytope $P=\text{conv}(A)$, for the matrices $A$ whose column vectors define the monomial parametrizations. Note that this approach can only be used to compute the generic ED degree of projective toric varieties.

\paragraph{Numerical methods}

In this paper, we also aimed to test how far the computations of these invariants for phylogenetic varieties could be extended using numerical methods. In this case, we turn the computation of critical points of the optimization problems \eqref{eq:du} and \eqref{eq:gen_du} into the problem of solving the systems of polynomial equations \eqref{eq:critEqsED} and \eqref{eq:critEqsgED}. We solve these polynomial systems via homotopy continuation and monodromy and use the implementations in the Julia package \texttt{HomotopyContinuation.jl} \cite{HC}. 

 In order to solve the systems using \emph{homotopy continuation}, a starting system with easily computed solutions is first derived using a polyhedral homotopy algorithm developed by Hubert and Sturmfels \cite{HuSt95}. After that, the original system is subjected to a parameter homotopy, which along with the solutions of the starting system defines the solution path that we can track.

As an alternative approach, we use \emph{monodromy} to compute all the isolated solutions with the specified parameter values and initial solutions. 
In order to use monodromy one must compute a starting solution in terms of variables $\theta_0$ and parameters $u_0$. 
One approach to do this is to substitute some random variables $\theta_0$ into the system \eqref{eq:critEqs} and then find initial parameters $u_0$ by solving the resulting linear system. 
However, in our systems, the number of parameters is larger than the number of variables and therefore, we get an underdetermined linear system. To solve it we assign random numbers to entries of $u_0$ until we get a full rank square linear system which we then solve for the remaining entries of $u_0$.
After finding a pair of data $u_0$ and parameters $\theta_0$ satisfying the system \eqref{eq:critEqs} the new solutions for $\theta$ are found by making loops in the parameter space of the system.

Both \emph{homotopy continuation} and \emph{monodromy} can be used straightforward to solve the systems \eqref{eq:critEqsED} and \eqref{eq:critEqsgED}. However, the computation of the ML degree has an extra difficulty, as the system of critical equations given in \eqref{eq:critEqs} is rational. However, it can be converted into a polynomial system by adding new variables as suggested in \cite[Section 3.3]{ASCB2005}. In this way, one can also use \emph{homotopy continuation} methods. However, for our specific models, solving the corresponding systems became computationally infeasible as the number of leaves and parameters in the model increased. This is mostly due to the fact that the number of paths to track increases at a significantly higher rate.
As a result, most computations did not finish due to time and memory constraints. We also use \emph{monodromy} to directly solve the rational system \eqref{eq:critEqs}.

It is important to emphasize that when computing the MLdegrees, we do not consider varieties after the Fourier coordinates change. Instead, we are calculating the MLdegree of the varieties with probability coordinates, which is why these degrees may increase compared to the degree of the original variety. Moreover, during the Julia computations, no correlation was observed between the volume of the associated polytope and the ML degrees.

Finally, we use a method based on interval arithmetic to certify solutions of the polynomial systems above obtained via numerical methods, see \cite{breiding2021certifying}. The method \texttt{certify} in the package \texttt{HomotopyContinuation.jl} uses interval arithmetic to prove the correctness of an isolated nonsingular solution to a (square) system of polynomial equations.
Given a numerical approximation of a complex zero, this package computes intervals for the real and imaginary parts of each variable that contain a true solution to the system and certifies whether the proposed numerical solution lies in this interval. This process may not be able to certify all true solutions. Thus the number of certified numerical solutions to our system of equations is a lower bound for the actual algebraic degrees of the corresponding varieties, see \cite[Section 1.1]{breiding2021certifying}.

Once we have obtained and certified the solutions for our systems we discard the ones providing singular points (see Remark \ref{rk:singular}), that is, the solutions with some zero entry. Finally, we find the number of unique points in the variety defined by our solutions by dividing the number of solutions by the cardinality of the fiber of the corresponding parametrization map (see Lemma \ref{lema:cardinallity}).
This allows us to turn our numerical computations into rigorous proof of our results in the following subsection.

\subsection{Algebraic degrees of phylogenetic varieties for small trees}\label{sec:degree-computations}

We display the results of our computations of the algebraic degrees of the corresponding phylogenetic varieties in Table \ref{table:3leaves}, Table \ref{table:4leaves}, and Table \ref{table:5leaves}, and the corresponding methods used to compute each of those degrees.
We also present the degree and dimension of the projective variety $\overline{\VV_T^\MM}$. The latter invariant corresponds to the number of free parameters needed to parameterize the variety plus one. Observe that, by intersecting the corresponding variety with the hyperplane given by $\left\{q_{\tt A \dots \tt A} = 1\right\}$, the dimension drops by one giving the dimension of ${\VV_T^\MM}$ which is the minimum number of parameters needed to determine the model, see \cite{CGS}.

\paragraph{3-leaf trees:}
We start by considering the binary $3$-leaf tree.

\begin{table}[H]
\centering
\def\arraystretch{1.5}
\setlength{\tabcolsep}{6pt}
\tikzset{every picture/.style={line width=0.75pt}} 
\begin{threeparttable}
\begin{tabular}{|M{1.7cm}|M{1cm}||M{0.8cm}|M{0.8cm}||M{1.3cm}||M{1.7cm}|M{1.7cm}|M{1.7cm}|M{1.7cm}|}
\hline
Tree & Model & dim & deg & ML deg & ED deg of ${\VV_T^\MM}$ & gED deg of ${\VV_T^\MM}$ & ED deg of $\overline{\VV_T^\MM}$ & gED deg of $\overline{\VV_T^\MM}$ \\ \hline \hline
 \multirow{4}{*}{ 
\begin{tikzpicture}[x=0.75pt,y=0.75pt,yscale=-0.4,xscale=0.4]
\draw [line width=1.2] (70.2,148.82) -- (120,209.02) ;
\draw [line width=1.2] (120,209.02) -- (70.6,269.82) ;
\draw [line width=1.2] (120,209.02) -- (195,209.02) ;
\end{tikzpicture}} 
 & CFN & $4$ & $1$ & $1$\tnotex{tn:hc} \tnotex{tn:mon}  & $1$\tnotex{tn:hc} \tnotex{tn:mon} \tnotex{tn:msolve} & $1$\tnotex{tn:hc} \tnotex{tn:mon} \tnotex{tn:msolve} & $1$\tnotex{tn:hc} \tnotex{tn:mon} \tnotex{tn:msolve} & $1$\tnotex{tn:hc} \tnotex{tn:mon} \tnotex{tn:msolve} \tnotex{tn:M2} \\ \cline{2-9} 
 & JC69 & $4$ & $3$ & $23$\tnotex{tn:hc} \tnotex{tn:mon}  & $13$\tnotex{tn:hc} \tnotex{tn:mon} \tnotex{tn:msolve} & $13$\tnotex{tn:hc} \tnotex{tn:mon} \tnotex{tn:msolve} & $18$\tnotex{tn:hc} \tnotex{tn:mon} \tnotex{tn:msolve} & $18$\tnotex{tn:hc} \tnotex{tn:mon} \tnotex{tn:msolve} \tnotex{tn:M2}\\ \cline{2-9} 
 & K80 & $7$ & $12$ & 1964\tnotex{tn:hc} \tnotex{tn:mon} & $184$\tnotex{tn:hc} \tnotex{tn:mon} \tnotex{tn:msolve}& $184$\tnotex{tn:hc} \tnotex{tn:mon} \tnotex{tn:msolve} & $188$\tnotex{tn:hc} \tnotex{tn:mon} \tnotex{tn:msolve} & $236$\tnotex{tn:hc} \tnotex{tn:mon} \tnotex{tn:msolve} \tnotex{tn:M2}\\ \cline{2-9} 
 & K81 & $10$ & $96$ & $-$ & $12673$\tnotex{tn:hc} \tnotex{tn:mon} & $15553$\tnotex{tn:hc} \tnotex{tn:mon} & $10360$\tnotex{tn:hc} \tnotex{tn:mon} & $21112$\tnotex{tn:hc} \tnotex{tn:mon}\\ \hline
\end{tabular}
\begin{tablenotes}[para,raggedright]
\item \textbf{Methods:}
\item[$\star$]\label{tn:hc}\emph{homotopy continuation}
\item[$\circ$] \label{tn:mon}\emph{monodromy}
\item[$\triangleright$] \label{tn:msolve}\emph{Gröbner basis}
\item[$\diamond$] \label{tn:M2}\emph{toricInvariants}
\end{tablenotes}
\caption{\label{table:3leaves} Dimension, degree, ML degree of ${\VV_T^\MM}$, ED and gED degrees of ${\VV_T^\MM}$ and $\overline{\VV_T^\MM}$ for the binary $3$-leaf tree illustrated in the first column and the four group-based models presented in Section \ref{sec:models}.}
\end{threeparttable}
\end{table}

\tnotex{tn:hc} \tnotex{tn:mon} \tnotex{tn:msolve}

\paragraph{4-leaf trees:} In the case of phylogenetic trees with four leaves, we consider both the star tree and the binary tree.
\begin{table}[H]
\centering
\def\arraystretch{1.6}
\setlength{\tabcolsep}{6pt}
\tikzset{every picture/.style={line width=0.75pt}} 
\begin{threeparttable}
\begin{tabular}{|M{1.7cm}|M{1cm}||M{0.8cm}|M{0.8cm}||M{1.3cm}||M{1.7cm}|M{1.7cm}|M{1.7cm}|M{1.7cm}|}
\hline
Tree & Model & dim & deg & ML deg & ED deg of ${\VV_T^\MM}$ & gED deg of ${\VV_T^\MM}$ & ED deg of $\overline{\VV_T^\MM}$ & gED deg of $\overline{\VV_T^\MM}$ \\ \hline \hline
\multirow{3}{*}{\begin{tikzpicture}[x=0.75pt,y=0.75pt,yscale=-0.4,xscale=0.4]
\draw [line width=1.2] (70.2,148.82) -- (120,209.02) ;
\draw [line width=1.2] (120,209.02) -- (70.6,269.82) ;
\draw [line width=1.2] (170.56,268.99) -- (120.5,209.02) ;
\draw [line width=1.2] (120.5,209.02) -- (169.63,148) ;
\end{tikzpicture}} 
& CFN & $5$ & $8$ & $92$\tnotex{tn:hc} \tnotex{tn:mon}  & $92$\tnotex{tn:hc} \tnotex{tn:mon} \tnotex{tn:msolve} & $92$\tnotex{tn:hc} \tnotex{tn:mon} \tnotex{tn:msolve} & $72$\tnotex{tn:hc} \tnotex{tn:mon} \tnotex{tn:msolve} & $120$\tnotex{tn:hc} \tnotex{tn:mon} \tnotex{tn:msolve} \tnotex{tn:M2}\\ \cline{2-9} 
 & JC69 & $5$ & $20$ & $4315$\tnotex{tn:hc} \tnotex{tn:mon}  & $124$\tnotex{tn:hc} \tnotex{tn:mon} \tnotex{tn:msolve} & $220$\tnotex{tn:hc} \tnotex{tn:mon} \tnotex{tn:msolve} & $243$\tnotex{tn:hc} \tnotex{tn:mon} \tnotex{tn:msolve} & $267$\tnotex{tn:hc} \tnotex{tn:mon} \tnotex{tn:msolve} \tnotex{tn:M2}\\ \cline{2-9} 
 & K80 & $9$ & $832$ & $-$ & $50972$\tnotex{tn:hc} \tnotex{tn:mon} & $88076$\tnotex{tn:hc} \tnotex{tn:mon} & $33176$\tnotex{tn:hc} \tnotex{tn:mon} & $106744$\tnotex{tn:hc} \tnotex{tn:mon}\\ \cline{2-9} 
\hhline{|=|=||=|=||=||=|=|=|=}

\multirow{2}{*}{\begin{tikzpicture}[x=0.75pt,y=0.75pt,yscale=-0.34,xscale=0.34]
\draw [line width=1.2] (70.2,148.82) -- (120,209.02) ;
\draw [line width=1.2] (120,209.02) -- (70.6,269.82) ;
\draw [line width=1.2] (120,209.02) -- (195,209.02) ;
\draw [line width=1.2] (245.06,268.99) -- (195,209.02) ;
\draw [line width=1.2] (195,209.02) -- (244.13,148) ;
\end{tikzpicture}} 
& CFN & $6$ & $4$ & $14$\tnotex{tn:hc} \tnotex{tn:mon}  & $10$\tnotex{tn:hc} \tnotex{tn:mon} \tnotex{tn:msolve}  & $30$\tnotex{tn:hc} \tnotex{tn:mon} \tnotex{tn:msolve}  & $4$\tnotex{tn:hc} \tnotex{tn:mon} \tnotex{tn:msolve}  & $36$\tnotex{tn:hc} \tnotex{tn:mon} \tnotex{tn:msolve} \tnotex{tn:M2} \\ \cline{2-9} 
 & JC69 & $6$ & $34$ & $17332$\tnotex{tn:mon}  & $290$\tnotex{tn:hc} \tnotex{tn:mon} \tnotex{tn:msolve}  & $630$\tnotex{tn:hc} \tnotex{tn:mon} \tnotex{tn:msolve}  & $309$\tnotex{tn:hc} \tnotex{tn:mon} \tnotex{tn:msolve}  & $809$\tnotex{tn:hc} \tnotex{tn:mon} \tnotex{tn:msolve} \tnotex{tn:M2} \\ \cline{2-9} 
\hline
\end{tabular}
\begin{tablenotes}[para,raggedright]
\item \textbf{Methods:}
\item[$\star$]\label{tn:hc}\emph{homotopy continuation}
\item[$\circ$] \label{tn:mon}\emph{monodromy}
\item[$\triangleright$] \label{tn:msolve}\emph{Gröbner basis}
\item[$\diamond$] \label{tn:M2}\emph{toricInvariants}
\end{tablenotes}
\caption{\label{table:4leaves} Dimension, degree, ML degree of ${\VV_T^\MM}$, ED and gED degrees of ${\VV_T^\MM}$ and $\overline{\VV_T^\MM}$ for the star $4$-leaf tree evolving under the CFN, JC69 and K80 models and also for the binary $4$-leaf tree with the CFN and JC69 models.}
\end{threeparttable}
\end{table}

\paragraph{5-leaf trees:}
Finally, we consider three different $5$-leaf trees, the star tree, an unresolved tree, and the binary tree.

\begin{table}[H]
\centering
\def\arraystretch{1.6}
\setlength{\tabcolsep}{6pt}
\tikzset{every picture/.style={line width=0.75pt}} 
\begin{threeparttable}
\begin{tabular}{|M{1.7cm}|M{1cm}||M{0.8cm}|M{0.8cm}||M{1.3cm}||M{1.7cm}|M{1.7cm}|M{1.7cm}|M{1.7cm}|}
\hline
Tree & Model & dim & deg & ML deg & ED deg of ${\VV_T^\MM}$ & gED deg of ${\VV_T^\MM}$ & ED deg of $\overline{\VV_T^\MM}$ & gED deg of $\overline{\VV_T^\MM}$ \\ \hline \hline \multirow{2}{*}{\begin{tikzpicture}[x=0.75pt,y=0.75pt,yscale=-0.34,xscale=0.34]
\draw [line width=1.2] (70.2,148.82) -- (120,209.02) ;
\draw [line width=1.2] (120,209.02) -- (70.6,269.82) ;
\draw [line width=1.2] (170.56,268.99) -- (120.5,209.02) ;
\draw [line width=1.2] (120.5,209.02) -- (169.63,148) ;
\draw [line width=1.2] (120,209.02) -- (120,144.02) ;
\end{tikzpicture}
} 
& CFN & $6$ & $52$ & $4698$\tnotex{tn:mon} & $ 858$\tnotex{tn:hc} \tnotex{tn:mon} \tnotex{tn:msolve} & $1098$\tnotex{tn:hc} \tnotex{tn:mon} \tnotex{tn:msolve} & $484$\tnotex{tn:hc} \tnotex{tn:mon} \tnotex{tn:msolve} & $1364$\tnotex{tn:hc} \tnotex{tn:mon} \tnotex{tn:msolve} \tnotex{tn:M2} \\ \cline{2-9} 
 & JC69 & $6$ & $115$ & $-$ & $713$\tnotex{tn:hc} \tnotex{tn:mon} \tnotex{tn:msolve} & $ 2313$\tnotex{tn:hc} \tnotex{tn:mon} \tnotex{tn:msolve} & $1792$\tnotex{tn:hc} \tnotex{tn:mon} \tnotex{tn:msolve} & $ 2792$\tnotex{tn:hc} \tnotex{tn:mon} \tnotex{tn:M2}\\ \hhline{|=|=||=|=||=||=|=|=|=}
 \multirow{2}{*}{\begin{tikzpicture}[x=0.75pt,y=0.75pt,yscale=-0.34,xscale=0.34]
\draw [line width=1.2] (70.2,148.82) -- (120,209.02) ;
\draw [line width=1.2] (120,209.02) -- (70.6,269.82) ;
\draw [line width=1.2] (120,209.02) -- (195,209.02) ;
\draw [line width=1.2] (245.06,268.99) -- (195,209.02) ;
\draw [line width=1.2] (195,209.02) -- (244.13,148) ;
\draw [line width=1.2] (195,209.02) -- (250.5,209.02) ;
\end{tikzpicture}} 
& CFN & $7 $ & $ 44$ & $2098$\tnotex{tn:mon} & $334$\tnotex{tn:hc} \tnotex{tn:mon} \tnotex{tn:msolve} & $1026$\tnotex{tn:hc} \tnotex{tn:mon} \tnotex{tn:msolve}& $ 176$\tnotex{tn:hc} \tnotex{tn:mon} \tnotex{tn:msolve}&  $1256$\tnotex{tn:hc} \tnotex{tn:mon} \tnotex{tn:M2}\\ \cline{2-9} 
 & JC69 & $7 $ & $ 315$ & $-$ & $3819$\tnotex{tn:hc} \tnotex{tn:mon} \tnotex{tn:msolve} & $12787$\tnotex{tn:hc} \tnotex{tn:mon}& $4569$\tnotex{tn:hc} \tnotex{tn:mon} \tnotex{tn:msolve}  & $15977$\tnotex{tn:hc} \tnotex{tn:mon}  \tnotex{tn:M2} \\ \hhline{|=|=||=|=||=||=|=|=|=}
 \multirow{2}{*}{\begin{tikzpicture}[x=0.75pt,y=0.75pt,yscale=-0.34,xscale=0.34]
\draw [line width=1.2] (70.2,148.82) -- (120,209.02) ;
\draw [line width=1.2] (120,209.02) -- (70.6,269.82) ;
\draw [line width=1.2] (120,209.02) -- (195,209.02) ;
\draw [line width=1.2] (245.06,268.99) -- (195,209.02) ;
\draw [line width=1.2] (195,209.02) -- (244.13,148) ;
\draw [line width=1.2] (159,209.02) -- (159.5,144.52) ;
\end{tikzpicture}} 
& CFN & $8$ & $34$ & $668$\tnotex{tn:mon} & $144$\tnotex{tn:hc} \tnotex{tn:mon} \tnotex{tn:msolve} & $1044$\tnotex{tn:hc} \tnotex{tn:mon} \tnotex{tn:msolve} & $ 62$\tnotex{tn:hc} \tnotex{tn:mon} \tnotex{tn:msolve} & $1294$\tnotex{tn:hc} \tnotex{tn:mon} \tnotex{tn:M2} \\ \cline{2-9} 
 & JC69 & $8$ & $813$ & $-$ & $ 10073$\tnotex{tn:hc} \tnotex{tn:mon} \tnotex{tn:msolve} & $ 62057$\tnotex{tn:hc} \tnotex{tn:mon}  & $9011$\tnotex{tn:hc} \tnotex{tn:mon} \tnotex{tn:msolve}& $78583$\tnotex{tn:M2} \\ \hline
\end{tabular}
\begin{tablenotes}[para,raggedright]
\item \textbf{Methods:}
\item[$\star$]\label{tn:hc}\emph{homotopy continuation}
\item[$\circ$] \label{tn:mon}\emph{monodromy}
\item[$\triangleright$] \label{tn:msolve}\emph{Gröbner basis}
\item[$\diamond$] \label{tn:M2}\emph{toricInvariants}
\end{tablenotes}
\caption{\label{table:5leaves} Dimension, degree, ML degree of ${\VV_T^\MM}$, ED and gED degrees of ${\VV_T^\MM}$ and $\overline{\VV_T^\MM}$ for the star $5$-leaf tree, an unresolved $5$-leaf tree and the binary $5$-leaf tree evolving under the CFN and the JC69 model.}
\end{threeparttable}
\end{table}

\newpage
For computing the degrees presented in the tables \ref{table:3leaves}, \ref{table:4leaves} and \ref{table:5leaves} using numerical methods we have found and certified the numerical solutions of the systems \eqref{eq:critEqs}, \eqref{eq:critEqsED} and \eqref{eq:critEqsgED} which allows us to state the following result (see \cite{breiding2021certifying}):
\begin{thm}\label{main}
 The results presented in Tables \ref{table:3leaves}, \ref{table:4leaves}, and \ref{table:5leaves}, except for the gED degree of the projective variety for the binary $5$-leaf tree under the JC69 model, are lower bounds for the corresponding algebraic degrees of the respective phylogenetic varieties.
\end{thm}

Furthermore, as symbolic and numerical methods consistently lead to the same results, we confidently pose the following conjecture.
\begin{conjecture}\label{conj:main}
 The Euclidean distance degrees presented in Tables \ref{table:3leaves}, \ref{table:4leaves}, and \ref{table:5leaves}, with the exceptions for the star trees with $3$ and $4$ leaves under the K81 and K80 models respectively; the gED degree of the affine variety for the unresolved $5$-leaf tree under the JC69 model; and both gED degrees for the binary $5$-leaf trees under the JC69 model,
 are the corresponding algebraic degrees of the respective phylogenetic varieties.
\end{conjecture}

\paragraph{Execution time.}
The results presented in this section were computed on an iMac with the Apple M1 chip and 16 GB of memory running macOS Monterey 12.6, Julia version 1.8.1 and Macaulay2 version 1.20. We used versions 2.6.4, 3.01, and 0.13.0 of the packages \texttt{HomotopyContinuation.jl} and \texttt{toricInvariants} and \texttt{Oscar} respectively.
In Table \ref{table:timings} we display the times (in seconds) needed to compute the ML and ED degrees of ${\VV_T^\MM}$ for all small trees and models considered above. We report timings using both numerical methods homotopy continuation and monodromy and the symbolic approach using msolve methods and discuss the results in the next section. We present the mean and standard deviation of the times needed to compute the corresponding degree ten times. For each iteration, we take a random data point $u\in\RR^m$ and a random prime $p$ in the case of symbolic computations. The blanks in the table correspond to cases that we have not been able to compute, see Section \ref{sec:discussion}.

\begin{table}[h]
\centering
\def\arraystretch{1.5}
\setlength{\tabcolsep}{3pt}
 \makebox[\linewidth]{
\begin{tabular}{|M{1.3cm}|M{1cm}||M{2.5cm}|M{3.5cm}||M{2.5cm}|M{3cm}|M{3cm}|}
\hline
\multirow{3}{*}{Tree} & \multirow{3}{*}{Model} & \multicolumn{2}{c||}{ML degree}                        & \multicolumn{3}{c|}{ED degree of ${\VV_T^\MM}$}                                                           \\ \cline{3-7} 
                      &                        & {\small homotopy continuation}                       & \multicolumn{1}{c||}{\small monodromy}                                             &{\small homotopy continuation}                               & \multicolumn{1}{c|}{\small monodromy}       & {\small Gröbner basis (msolve)} \\ \hline\hline
\multirow{4}{*}{\begin{tikzpicture}[x=0.75pt,y=0.75pt,yscale=-0.25,xscale=0.25] \draw [line width=1.2] (70.2,148.82) -- (120,209.02) ; \draw [line width=1.2] (120,209.02) -- (70.6,269.82) ; \draw [line width=1.2] (120,209.02) -- (195,209.02) ;\end{tikzpicture}}  
                      &     CFN                & \multicolumn{1}{c|}{\footnotesize $0.099 \pm 0.020 $}       & \multicolumn{1}{c||}{\footnotesize $0.047 \pm 0.003 $}        & \multicolumn{1}{c|}{\footnotesize $0.021 \pm 0.004$}     & \multicolumn{1}{c|}{\footnotesize $0.023 \pm 0.002$}  & \multicolumn{1}{c|}{\footnotesize $0.001 \pm 0.001$}    \\ \cline{2-7} 
                      &     JC69               & \multicolumn{1}{c|}{\footnotesize $0.366 \pm 0.056$}        & \multicolumn{1}{c||}{\footnotesize $0.473 \pm 0.035 $}        & \multicolumn{1}{c|}{\footnotesize $0.015 \pm 0.001$}     & \multicolumn{1}{c|}{\footnotesize $0.123 \pm 0.021$}  & \multicolumn{1}{c|}{\footnotesize $  0.001\pm 6\cdot10^{-6}$}       \\ \cline{2-7} 
                      &     K80                & \multicolumn{1}{c|}{\footnotesize $2125.988 \pm 47.931 $}   & \multicolumn{1}{c||}{\footnotesize$57.490 \pm 6.268 $}        & \multicolumn{1}{c|}{\footnotesize $3.518 \pm 0.269$}     & \multicolumn{1}{c|}{\footnotesize $1.248 \pm 0.212$}  & \multicolumn{1}{c|}{\footnotesize $ 1.43 \pm 0.086$}       \\ \cline{2-7} 
                      &     K81                & \multicolumn{1}{c|}{\footnotesize $-$}   & \multicolumn{1}{c||}{$-$}                                                        & \multicolumn{1}{c|}{\footnotesize $389.514 \pm 19.449$}   & \multicolumn{1}{c|}{\footnotesize $146.447\pm 11.589$}          &    \multicolumn{1}{c|}{$-$}    \\ \hline\hline
\multirow{3}{*}{\begin{tikzpicture}[x=0.75pt,y=0.75pt,yscale=-0.25,xscale=0.25]\draw [line width=1.2] (70.2,148.82) -- (120,209.02) ;\draw [line width=1.2] (120,209.02) -- (70.6,269.82) ;\draw [line width=1.2] (170.56,268.99) -- (120.5,209.02) ;\draw [line width=1.2] (120.5,209.02) -- (169.63,148) ;\end{tikzpicture}}     
                      &     CFN                & \multicolumn{1}{c|}{\footnotesize $30.032 \pm 1.096 $}      & \multicolumn{1}{c||}{\footnotesize $7.212 \pm 0.355 $}        & \multicolumn{1}{c|}{\footnotesize $0.106 \pm 0.009$}     & \multicolumn{1}{c|}{\footnotesize $3.736 \pm 0.455 $}  & \multicolumn{1}{c|}{\footnotesize $0.036 \pm 0.003$}       \\ \cline{2-7} 
                      &     JC69               & \multicolumn{1}{c|}{\footnotesize $544.109 \pm 8.975 $}     & \multicolumn{1}{c||}{\footnotesize $1722.446 \pm 1218.715 $}  & \multicolumn{1}{c|}{\footnotesize $0.140 \pm 0.006$}     & \multicolumn{1}{c|}{\footnotesize $1.835 \pm 0.082 $}  & \multicolumn{1}{c|}{\footnotesize $0.097 \pm 0.041$}       \\ \cline{2-7} 
                      &     K80                & \multicolumn{1}{c|}{$-$}                                    & \multicolumn{1}{c||}{$-$}                                     & \multicolumn{1}{c|}{\footnotesize $634.988 \pm 13.45$}   & \multicolumn{1}{c|}{\footnotesize $3808.368\pm 545.587$}          &   \multicolumn{1}{c|}{$-$}     \\ \hline
\multirow{2}{*}{\begin{tikzpicture}[x=0.75pt,y=0.75pt,yscale=-0.25,xscale=0.25]\draw [line width=1.2] (70.2,148.82) -- (120,209.02) ;\draw [line width=1.2] (120,209.02) -- (70.6,269.82) ;\draw [line width=1.2] (120,209.02) -- (195,209.02) ;\draw [line width=1.2] (245.06,268.99) -- (195,209.02) ;\draw [line width=1.2] (195,209.02) -- (244.13,148) ;\end{tikzpicture}}     
                      &     CFN                & \multicolumn{1}{c|}{\footnotesize $434.087 \pm 10.165 $}    & \multicolumn{1}{c||}{\footnotesize $2.537 \pm 0.756$}         & \multicolumn{1}{c|}{\footnotesize $2.032 \pm 0.228$}     & \multicolumn{1}{c|}{\footnotesize $0.652 \pm 0.107 $}  & \multicolumn{1}{c|}{\footnotesize $ 0.004\pm 0.0001$}       \\ \cline{2-7} 
                      &     JC69               & \multicolumn{1}{c|}{$-$}                                    & \multicolumn{1}{c||}{\footnotesize $6226.114 \pm 3517.122$}   & \multicolumn{1}{c|}{\footnotesize $1.875 \pm 0.112$}     & \multicolumn{1}{c|}{\footnotesize $4.692 \pm 0.953 $}  & \multicolumn{1}{c|}{\footnotesize $ 1.676\pm 0.068$}       \\ \hline\hline
\multirow{2}{*}{\begin{tikzpicture}[x=0.75pt,y=0.75pt,yscale=-0.25,xscale=0.25]\draw [line width=1.2] (70.2,148.82) -- (120,209.02) ;\draw [line width=1.2] (120,209.02) -- (70.6,269.82) ;\draw [line width=1.2] (170.56,268.99) -- (120.5,209.02) ;\draw [line width=1.2] (120.5,209.02) -- (169.63,148) ;\draw [line width=1.2] (120,209.02) -- (120,144.02) ;\end{tikzpicture}}     
                      &     CFN                & \multicolumn{1}{c|}{$-$}                                    & \multicolumn{1}{c||}{\footnotesize $204.914 \pm 17.318 $}     & \multicolumn{1}{c|}{\footnotesize $1.787 \pm 0.049$}     & \multicolumn{1}{c|}{\footnotesize $10.623 \pm 1.626$} & \multicolumn{1}{c|}{\footnotesize $ 19.751\pm 1.814$}       \\ \cline{2-7} 
                      &     JC69               & \multicolumn{1}{c|}{$-$}                                    & \multicolumn{1}{c||}{$-$}                                     & \multicolumn{1}{c|}{\footnotesize $3.206 \pm 0.128$}     & \multicolumn{1}{c|}{\footnotesize $4.055 \pm 0.534$}  & \multicolumn{1}{c|}{\footnotesize $ 10.829\pm 0.693$}       \\ \hline
\multirow{2}{*}{\begin{tikzpicture}[x=0.75pt,y=0.75pt,yscale=-0.25,xscale=0.25]\draw [line width=1.2] (70.2,148.82) -- (120,209.02) ;\draw [line width=1.2] (120,209.02) -- (70.6,269.82) ;\draw [line width=1.2] (120,209.02) -- (195,209.02) ;\draw [line width=1.2] (245.06,268.99) -- (195,209.02) ;\draw [line width=1.2] (195,209.02) -- (244.13,148) ;\draw [line width=1.2] (195,209.02) -- (250.5,209.02) ;\end{tikzpicture}}     
                      &     CFN                & \multicolumn{1}{c|}{$-$}                                    & \multicolumn{1}{c||}{\footnotesize $479.032 \pm 93.423 $}     & \multicolumn{1}{c|}{\footnotesize $39.334 \pm 1.597$}    & \multicolumn{1}{c|}{\footnotesize $24.206 \pm 6.310 $}  & \multicolumn{1}{c|}{\footnotesize $ 18.260\pm 0.303$}       \\ \cline{2-7} 
                      &     JC69               & \multicolumn{1}{c|}{$-$}                                    & \multicolumn{1}{c||}{$-$}                                     & \multicolumn{1}{c|}{\footnotesize $55.582 \pm 1.592$}    & \multicolumn{1}{c|}{\footnotesize $65.261 \pm 13.539 $} & \multicolumn{1}{c|}{\footnotesize $ 3899.185 \pm 1598.814$}       \\ \hline
\multirow{2}{*}{\begin{tikzpicture}[x=0.75pt,y=0.75pt,yscale=-0.25,xscale=0.25]\draw [line width=1.2] (70.2,148.82) -- (120,209.02) ;\draw [line width=1.2] (120,209.02) -- (70.6,269.82) ;\draw [line width=1.2] (120,209.02) -- (195,209.02) ;\draw [line width=1.2] (245.06,268.99) -- (195,209.02) ;\draw [line width=1.2] (195,209.02) -- (244.13,148) ;\draw [line width=1.2] (159,209.02) -- (159.5,144.52) ;\end{tikzpicture}}     
                      &     CFN                & \multicolumn{1}{c|}{$-$}                                    & \multicolumn{1}{c||}{\footnotesize $995.213 \pm 159.482 $}    & \multicolumn{1}{c|}{\footnotesize $245.454 \pm 7.232$}   & \multicolumn{1}{c|}{\footnotesize $22.380 \pm 1.765 $}    & \multicolumn{1}{c|}{\footnotesize $ 16.551\pm 0.639$}       \\ \cline{2-7} 
                      &     JC69               & \multicolumn{1}{c|}{$-$}                                    & \multicolumn{1}{c||}{$-$}                                     & \multicolumn{1}{c|}{\footnotesize $810.829 \pm 12.682$}  & \multicolumn{1}{c|}{\footnotesize $ 855.017 \pm 166.026$} & \multicolumn{1}{c|}{\footnotesize $ 30671.873 \pm 4067.029$}       \\ \hline
\end{tabular}}
\caption{\label{table:timings} We present the mean and standard deviation of the computational times (in seconds) needed to obtain the ML and ED degrees of ${\VV_T^\MM}$ using homotopy continuation, monodromy and msolve for ten iterations. }
\end{table}

\section{Discussion}\label{sec:discussion}

The computations performed in this work allow us to compare the advantages and limitations of symbolic versus numerical methods. We observed that while numerical methods were able to compute some degrees that the symbolic methods did not, the symbolic approaches are faster for a small number of parameters.

The symbolic computations performed via computing Gröbner basis have been possible to perform on finite fields $\mathbb{Z}/p\mathbb{ZZ}$ and using the \texttt{msolve} library. The bottleneck of such computations is the computation of the saturation ideal. Computing such saturation over the rationals, by computing Gröbner basis on algebra software such as \texttt{Macaulay2} \cite{M2} or \texttt{Magma} \cite{magma}, can only be carried out for trees under the CFN model and some under the JC69 model ($3$ and $4$ leaves), with much longer computation times.

It is also worth mentioning that the approach described in \cite{Helmer_Sturmfels_2018} allowed us to compute all but two of the gED degrees of the corresponding projective varieties (the star trees with 3 and 4 leaves under the K81 and K80 models respectively, were the lone exceptions). These symbolic computations were performed using the \texttt{M2} package \texttt{ToricInvariants.m2} \cite{ToricInvariants}. This approach outperforms any other method in the case of a binary tree on 5 leaves under the JC69 model since we have not been able to verify numerically the generic ED degree for the projective variety of this model. The closest we have gotten is 78,577 certified solutions. The numerical methods always fail to find all the solutions in this example.

Although the different approaches considered allowed us to present new bounds for the Euclidean distance degrees of the phylogenetic varieties we considered, we have not yet been able to compute all the ML degrees.
Moreover, there is evidence that we have reached the boundary of problems that are tractable using the existing numerical techniques. 
These missing degrees, along with the degrees of larger trees and more general models, remain an open question and the next step in our studies. 

\newpage
It is also important to note the differences between homotopy continuation and monodromy. In general, for all missing degrees in our tables, homotopy continuation methods run out of memory during the computation of a starting system and monodromy methods have excessively long computation times. Furthermore, in our calculations, we observe that although for most cases monodromy is slower, it manages to complete the computation for some cases that homotopy continuation is unable to do, as can be seen in Table \ref{table:timings}.

Regarding the time of execution, it seems that the symbolic approach by computing Gr\"obner basis using \texttt{msolve} beats the numerical methods for simpler models; however, when the number of parameters increases the numerical methods are more effective. In particular, for more than $6$ parameters (with models of $4$ states), computations using \texttt{msolve} are slower or not able to finish due to the lack of memory or extremely long computation times. For instance, our machine has been running without output for more than a week for the star tree with 3 leaves under the K81 model and for more than two weeks for the star tree with 4 leaves under the K80 model. In the case of ML degrees, the missing values that we tried to compute with homotopy continuation (first column of Table \ref{table:timings}), correspond to computations that failed due to memory overflow while attempting to compute a start system for homotopy continuation. In the case of monodromy, the missing entries are not due to memory issues but prohibitively long running times. For example, the computation for the star tree with 3 leaves under the K81 model halted after running for 6 days. Using monodromy, the system had found 10,080,070 solutions and exited halfway through the certification process. Similarly, for the binary tree with 4 leaves under the K80 model, after 3 days of running time, using monodromy, the system computed 2,466,126 solutions. At that moment, we paused the process and certified 62,141 solutions. This also demonstrates how fast these degrees grow when we add more free parameters to the models.

The key advantage of the numerical methods versus methods based on Gr\"obner basis computations over a finite field is that one can compute all solutions, and in particular, check the real and positive solutions. In practice, in the context of phylogenetics, one wants to find real and positive solutions to the studied problems.
In our simulations, we also kept track of the number of real solutions to our systems and generally, the systems studied do not have real and stochastic solutions. However, on a few occasions and for some random initial data we have been able to find one or two stochastic solutions. The number of real solutions slightly increases when the random data point is real but the increase in the number of stochastic, i.e. biologically meaningful, solutions is rather limited. This means that, from a biological point of view, if one wants to find the point of the variety with stochastic parameters closest to a given data point, one should take into account the boundary of the variety region defined by distributions that are the image of stochastic parameters. This, therefore, adds considerable complexity to the problem of finding the tree and parameters that best explain a given data point.

Finally, it would be interesting to extend our computations to include small phylogenetic networks. Networks, are more general models that allow to consideration of more complex evolutionary processes such as hybridization and gene flow, see \cite{networks} for a nice introduction to the topic. Moreover, we believe that it is possible to design tailored homotopy continuation methods that may allow the computation of significantly lower bounds for some of the missing algebraic degrees and we propose to explore this direction further.

\section*{Acknowledgments}
The authors would like to thank Bernd Sturmfels for introducing the three of us and encouraging us to work on this project. They also wish to thank Paul Breiding, Sonja Petrovi\'c, and Kemal Rose for helpful discussions on this topic. The research of Elima Shehu was funded by the Deutsche Forschungsgemeinschaft (DFG, German Research Foundation), Projektnummer 445466444.

\bibliographystyle{abbrv}
\bibliography{biblio}

\end{document}